\documentclass[a4paper,11pt]{article}

\usepackage[top=18truemm,bottom=25truemm,left=25truemm,right=25truemm]{geometry} 
\usepackage{indentfirst} 

\makeatletter 
\def\section{\@startsection {section}{1}{\z@}{-3.5ex plus -1ex minus -.2ex}{2.3 ex plus .2ex}{\LARGE\bf}}
\makeatother
\usepackage{secdot} 

\makeatletter 
\def\subsection{\@startsection {subsection}{1}{\z@}{-3.5ex plus -1ex minus -.2ex}{2.3 ex plus .2ex}{\Large\it}}
\makeatother

\usepackage{amsmath, amssymb} 

\usepackage[dvipdfmx]{graphicx}
\usepackage{float} 
\graphicspath{{./}} 
\usepackage[labelsep=period,figurename=Fig. ,tablename=Table]{caption} 
\usepackage{subcaption} 
\usepackage{booktabs}
\usepackage{multirow} 
\usepackage{tabularx}
\newcolumntype{C}{>{\centering\arraybackslash}X}

\usepackage{cite}
\usepackage{url}

\usepackage{comment} 
\usepackage{color}
\usepackage{ascmac}
\usepackage{fancybox}

\def\vector#1{\mbox{\boldmath $#1$}}

\title{Characterization of the formation structure in team sports\footnote{Translated version of the paper: T. Narizuka and Y. Yamazaki, ``Characterization of the formation structure in team sports'', Proceedings of the Institute of Statistical Mathematics, Vol. 65, No.2, 299-307, 2017 (in Japanese).}}

\usepackage{authblk}
\author[1]{Takuma Narizuka\thanks{{\it E-mail address}: pararel@gmail.com (T. Narizuka).\\
}}
\author[2]{Yoshihiro Yamazaki}
\affil[1]{Department of Physics, Faculty of Science and Engineering, Chuo University, Bunkyo, Tokyo 112-8551, Japan}
\affil[2]{Department of Physics, School of Advanced Science and Engineering, Waseda University, Shinjuku, Tokyo 169-8555, Japan}
\date{}

\begin{document}
	\maketitle
\begin{abstract}
We propose a method to identify the formation structure in team sports based on Delaunay triangulation.
The adjacency matrix obtained from the Delaunay triangulation for each player is regarded as the formation pattern.
Our method allows time-series analysis and a quantitative comparison of formations.
A classification algorithm of formations is also proposed by combining our method with hierarchical clustering.
\end{abstract}

\baselineskip 18pt
\section{Introduction}
In competitive sports, unexpected interactions among players under some rules generate various player motions such as scoring, ball passing, marking an opponent player, keeping formation, and so on.
Especially in team sports, cooperation among team members is an essential strategy.
For example, each player attempts to keep a certain distance with other players, and a formation structure emerges.
The formation structure generates an effective area around players, allowing efficient ball-passing and player marking.
Taki et al. have quantified such an effective area as a ``dominant region'' in which a certain player can arrive prior to any other players \cite{Taki1996}.
Typically, the dominant region is expressed by a Voronoi region \cite{Okabe2000}, and its fundamental properties have been analyzed for football games \cite{Kim2004, Fonseca2012}.
More realistic definition of the dominant region is given by the ``motion model'' in which the Voronoi region is modified by velocity and acceleration of players, and it has been applied to real game analysis \cite{Taki2000, Fujimura2005, Nakanishi2010, Gudmundsson2014, Gudmundsson2017}.

Meanwhile, several studies have developed the method called ``role representation'', which characterizes player positions such as ``defender'' or ``forward'' in the formation \cite{Bialkowski2014a, Bialkowski2014b, Lucey2013, Lucey2014, Wei2013}.
The main idea of the role representation is that each player is not distinguished by an uniform number but by a relative position (role) assigned them at each frame of the data.
Bialkowski et al. have proposed the algorithm assigning roles for players based on heat maps \cite{Bialkowski2014b}.
Their method has been reported to be useful for the analysis of positional exchange of players and the detection of characteristic formation such as ``4-4-2''.
%
%

In this letter, we propose another method characterizing formation structures.
Since our method is based on Delaunay triangulation, a formation is expressed as a network and time-series analysis and quantitative comparison of formations are possible.

\section{Definition of formation structures by Delaunay triangulation}
In order to characterize the formation of a team, we focus on the adjacency relationship of Voronoi regions assigned to each player.
It can be described by a Delaunay triangulation.
The Delaunay triangulation has a network structure (we call ``Delaunay network'' hereafter) and is quantified by the adjacency matrix $ \vector{A} $ whose component is defined as follows:
\begin{align*}
	A_{ij} &= 
	\begin{cases}
		1 & \textrm{the Voronoi regions of players $ i $ and $ j $ are adjacent each other},\\ 
		0 & \textrm{otherwise}.
	\end{cases}
\end{align*}
We define the formation at time $ t $ by the adjacency matrix $ \vector{A}(t) $ of Delaunay network.
It is noted that we decide the Voronoi region of each player by considering the boundary of the field.

In order to measure dissimilarity between $ \vector{A}(t) $ and $ \vector{A}(t') $, distance $ D_{tt'} $ is introduced as follows:
\begin{align}
	D_{tt'} &= \sum_{i=1}^{N} \sum_{j=1}^{N} [A_{ij}(t) - A_{ij}(t')]^{2}.
	\label{eq:dissim}
\end{align}
Here, we define $ D_{tt'} $ as the Euclidean squared distance since we perform the hierarchical clustering by using Wards' method in the next section.
By using $ D_{tt'} $, comparison between two different formations is possible quantitatively.

\begin{figure}[H]
	\centering
	\includegraphics[width=8cm]{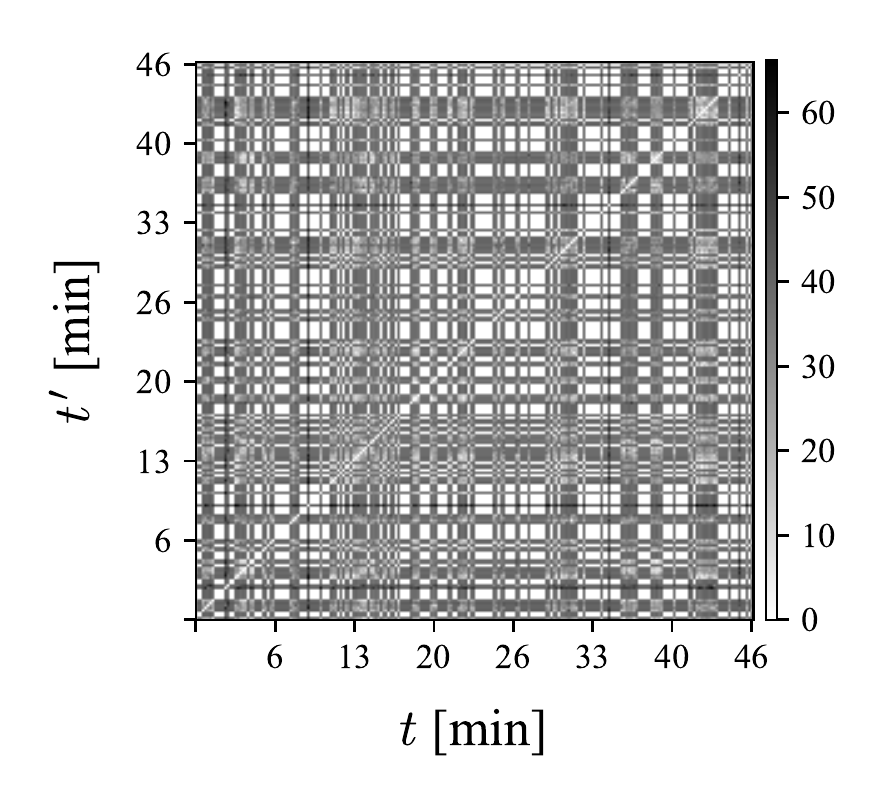}
	\caption{Visualization of dissimilarity matrix $ \vector{D} $. The value of $ D_{tt'} $ is expressed by grayscale.}
	\label{fig:Dmat}
\end{figure}

\begin{figure}[H]
	\centering
	\includegraphics[width=14cm]{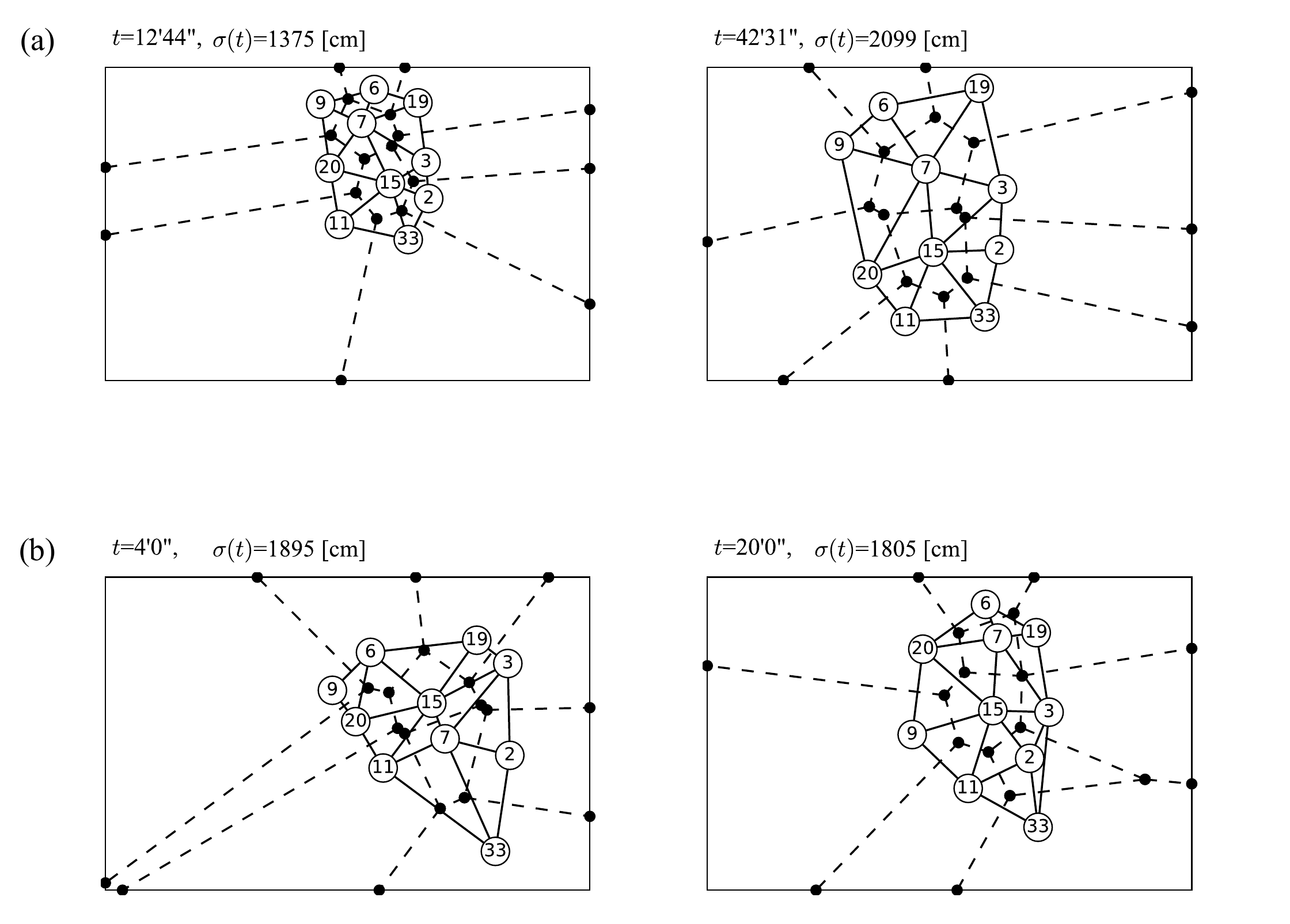}
	\caption{Comparison of two different formations. (a) $ D_{tt'}=0 $. (b) $ D_{tt'}=30 $. The gyration radius $ \sigma(t) $ is shown above each panel.}
	\label{fig:form_comparison}
\end{figure}

Here, we demonstrate our characterization of formation with real data obtained from football games.
In fact, player tracking data of football games, which includes all player positions every 0.04 seconds, provided by DataStadium Inc., Japan is used.
We focus on the J-League match between Iwata and Nagoya held on February 27, 2016, and analyze data for 10 players ($ N=10 $) except a goal keeper for each team.
Figure \ref{fig:Dmat} visualizes the dissimilarity matrix $ \vector{D} $ whose component corresponds to the value of $ D_{tt'} $.
In this figure, the region of small $ D_{tt'} $ is observed repeatedly.
It indicates that similar formations emerge frequently and intermittently in the game.
We also show two Delaunay networks with $ D_{tt'}=0 $ in Fig. \ref{fig:form_comparison}(a).
We clearly confirm that they have the same adjacency relationship.
Here, the deviation from the centroid of a team can be measured by the gyration radius given by
\begin{align*}
	\sigma(t) &= \sqrt{\frac{1}{N} \sum_{j=1}^{N}|\vector{x}_{c}(t) - \vector{x}_{j}(t)|^{2}},
\end{align*}
where $ \vector{x}_{c}(t) $ and $ \vector{x}_{j}(t) $ denote the centroid position of the team and $ j $-th player's position, respectively.
Even though the values of $ \sigma $ for each formation in Fig. \ref{fig:form_comparison}(a) are different, our method can extract a basic formation structure by players independently of $ \sigma $.
Figure \ref{fig:form_comparison}(b) is another example where $ D_{tt'}=30 $.
It is confirmed that the dissimilarity becomes large when players exchange their positions each other.


\section{Classification of formation by using hierarchical clustering}
\begin{figure}[H]
	\centering
	\includegraphics[width=13cm]{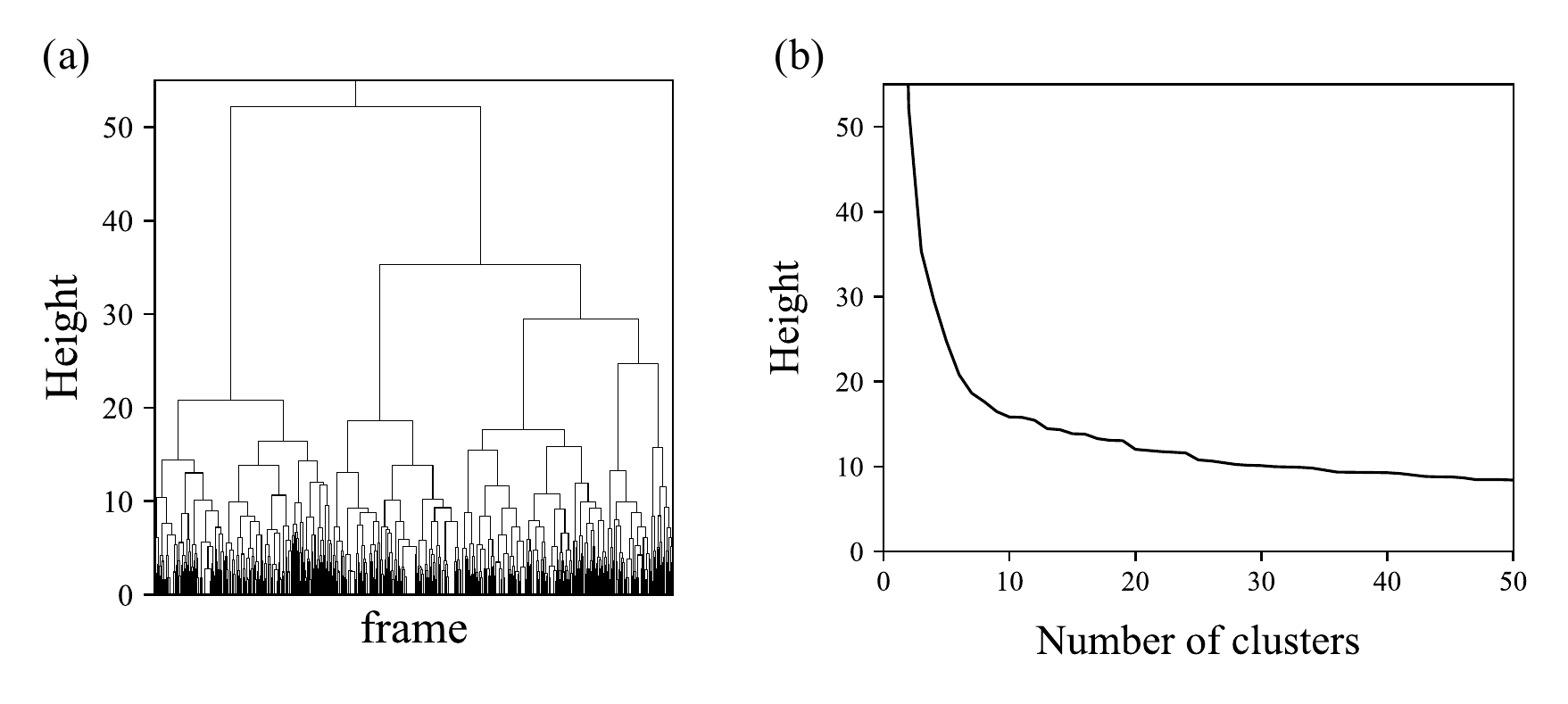}
	\caption{(a) Dendrogram and (b) the relation between the number of clusters and height obtained from the hierarchical clustering. The height of the vertical axis corresponds to the distance between two merged clusters. We choose $ h_{c} = 15 $ in this case.}
	\label{fig:dedr_scree}
\end{figure}

In this section, we propose the classification algorithm of Delaunay networks at different times in the following four steps (i)-(iv).
(i) We calculate Delaunay networks every frames.
%
%
(ii) Hierarchical clustering with Ward's method is executed by using the dissimilarity measure defined by eq. \eqref{eq:dissim}.
For the Ward's method, distance between two clusters $ C_{1} $ and $ C_{2} $ is defined as follows \cite{Tan2006}:
\begin{align*}
	h(C_{1}, C_{2}) &= V(C_{1} \cup C_{2}) - [V(C_{1}) + V(C_{2})],
\end{align*}
where $ V(C) $ denotes the sum of Euclidean squared distance between each point in the cluster $ C $ and the centroid of $ C $.
At the initial state where each cluster contains one Delaunay network, $ h(C_{1}, C_{2})=V(C_{1} \cup C_{2}) $ is equivalent to equation \eqref{eq:dissim}.
(iii) The result of the hierarchical clustering is displayed by the dendrogram whose vertical axis (height) corresponds to the distance between two merged clusters [Fig. \ref{fig:dedr_scree}(a)].
We extract $ N_{c} $ clusters by cutting the dendrogram at certain height $ h_{c} $.
The value of $ h_{c} $ is determined as the point where height increases rapidly with decreasing of the number of clusters [Fig. \ref{fig:dedr_scree}(b)].
In the case of Fig. \ref{fig:dedr_scree}, $ N_{c}=12 $ is obtained by choosing $ h_{c}=15 $.
(iv) Since each cluster contains similar Delaunay networks, we can visualize coarse grained formations from each cluster as follows.
For each Delaunay network in a cluster, we first convert the positional coordinate of $ j $-th player of a team, $ \vector{x}_{j}(t) $, to a normalized one $ \tilde{\vector{x}}_{j}(t) $ as
\begin{align*}
	\tilde{\vector{x}}_{j}(t) &= \frac{\vector{x}_{j}(t) - \vector{x}_{c}(t)}{\sigma(t)},
\end{align*}
where $ \vector{x}_{c}(t) $ and $ \sigma(t) $ denote the centroid position and the gyration radius of the team, respectively.
This transformation enables to compare each Delaunay network independently of $ \vector{x}_{c} $ and $ \sigma(t) $.
Next, we plot all Delaunay networks in the cluster, and visualize the time averaged position of each player by an ellipse.
It is noted that the direction and magnitude of the ellipse are determined by the eigen vector and eigen value of covariance matrix of each player's position.

\begin{figure}[H]
	\centering
	\includegraphics[width=13cm]{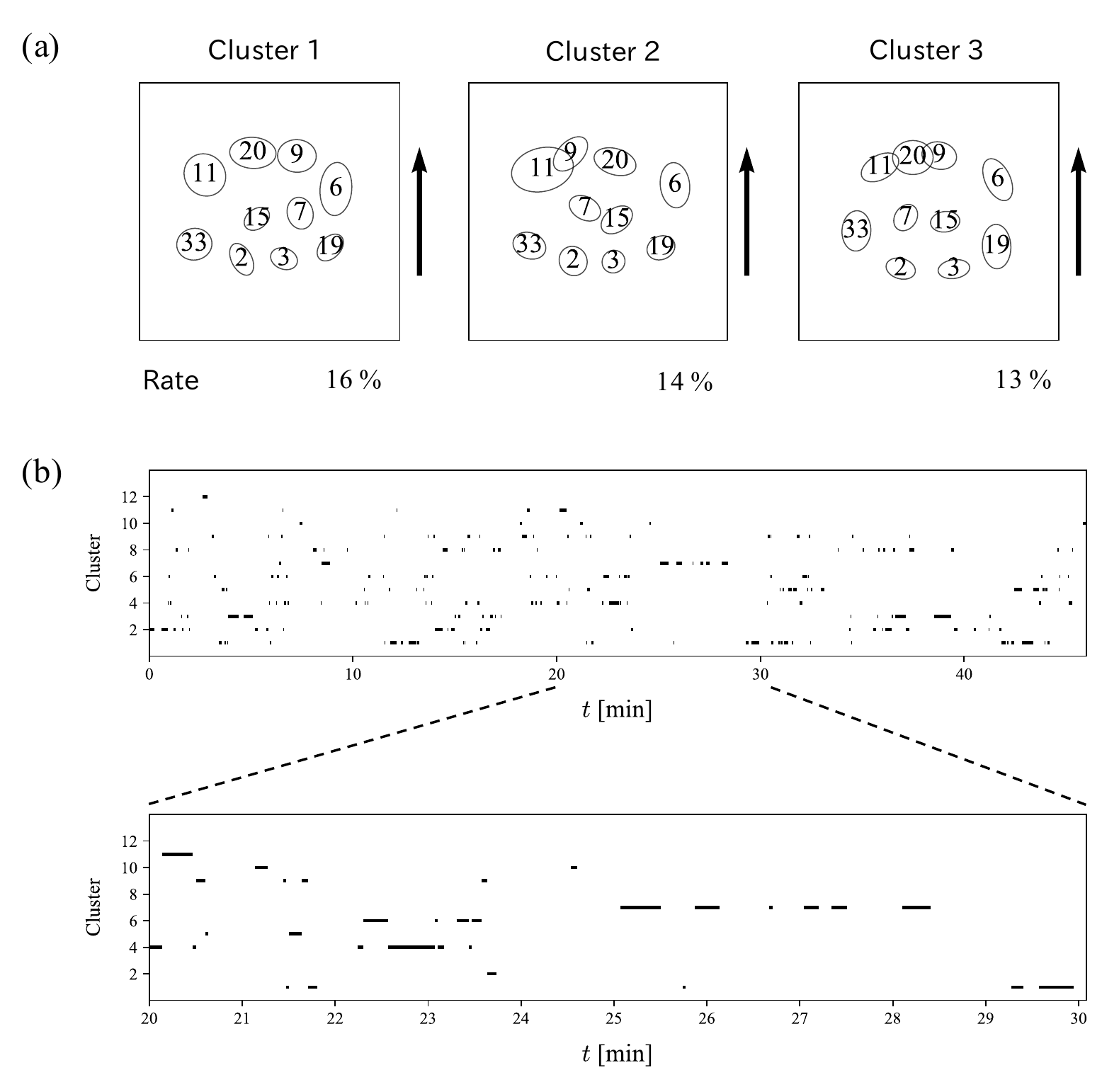}
	\caption{(a) Three largest clusters in $ N_{c}=12 $ clusters. (b) Time series of the clusters. The vertical axis indicates the cluster number. The enlarged view where $ 20 \leq t \leq 30 $ is shown in the below panel.}
	\label{fig:clus}
\end{figure}

Figure \ref{fig:clus}(a) is an example of coarse grained formations where $ h_{c}=15 $ and $ N_{c}=12 $.
Each panel in the figure corresponds to the three largest clusters (i.e., the most frequent formations in the game).
We find that several pairs of players exchange their positions each other: player 20 and 9, and player 7 and 15, for example.
We also show the time series of clusters during a game in Fig. \ref{fig:clus}(b); the vertical axis indicates the cluster number and the enlarged view where $ 20\leq t \leq 30 $ is shown in the below panel.
It is found that the formation of the team changes intermittently.

\begin{figure}[H]
	\centering
	\includegraphics[width=12cm]{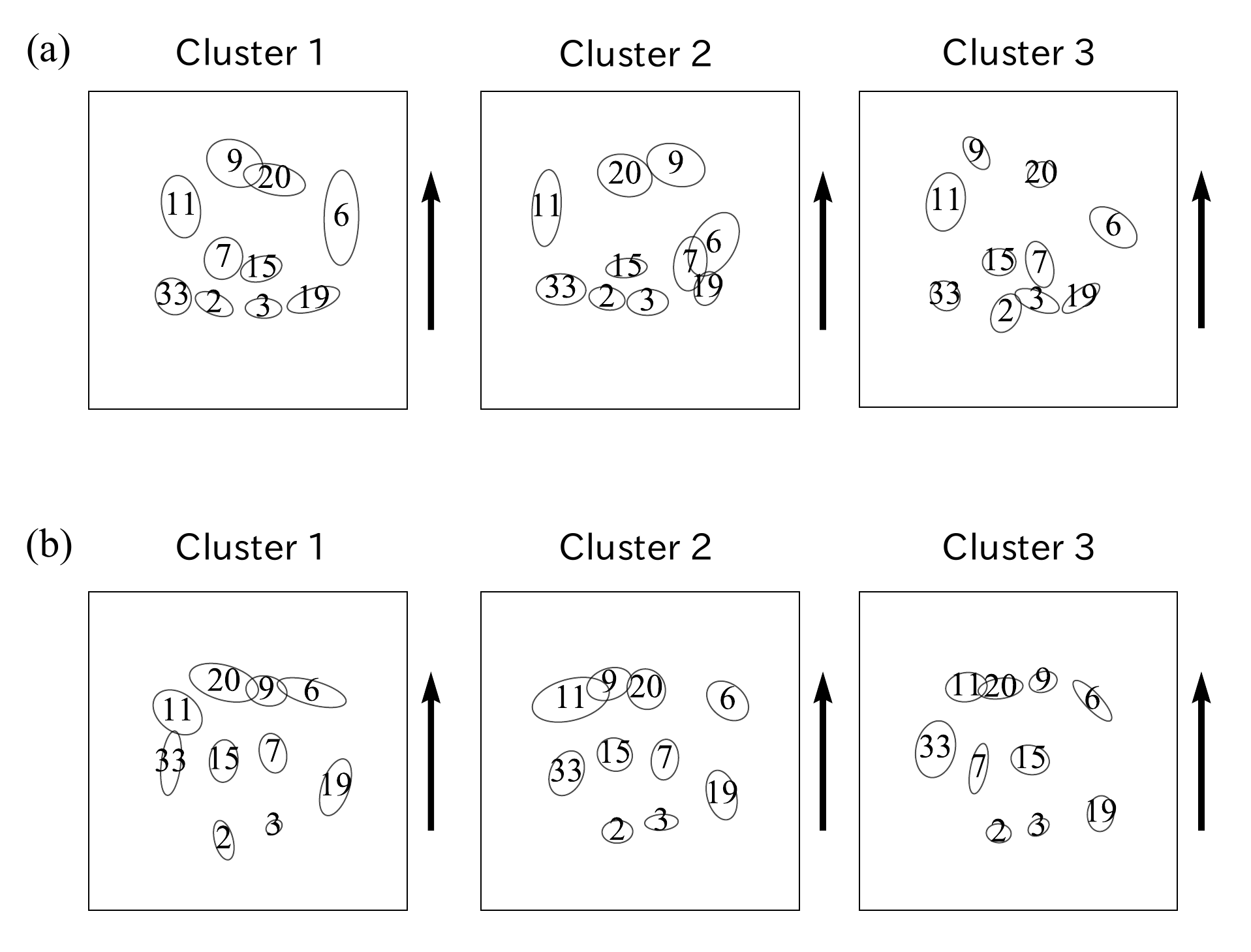}
	\caption{Results of clustering for the (a) offense and (b) defense scenes.}
	\label{fig:clus_of-df}
\end{figure}

As an application of our method, Delaunay networks in the offense and defense scenes can be distinguished separately.
The offense (defense) scene is defined as the scenes where the centroid position $ \vector{x}_{c}(t) $ is in one third from the opponent's (own) goal.
As shown in Fig. \ref{fig:clus_of-df}, different coarse grained formations such as ``4-4-2'' and ``2-4-4'' are observed in each scene.
Thus, our method can extract different formations depending on different situations.


\section{Summary and discussion}
We have proposed a new method for characterization of formation structures based on the Delaunay triangulation.
Our method defines a formation as the adjacency matrix of the Delaunay network and enables hierarchical clustering and the time-series analysis of formations.
We expect the following developments and applications of our method.

First, in eq. \eqref{eq:dissim}, players with many Delaunay edges (i.e., edges of the Delaunay network) increase $ D_{tt'} $ compared with other players.
Hence, more accurate clustering can be possible by extending the definition of dissimilarity $ D_{tt'} $ to
\begin{align*}
	D_{tt'} &= \sum_{i=1}^{N} \sum_{j=1}^{N} \left[ \frac{A_{ij}(t)}{k_{i}(t)} - \frac{A_{ij}(t')}{k_{i}(t')} \right]^{2},
\end{align*}
where $ k_{i}(t) $ represents the number of Delaunay edges of player $ i $.
Second, with respect to the game analysis based on the Delaunay network, correlation analysis of formations such as the transition among different clusters or interaction of between two teams can be considered.
And finally, it is also important to characterize the dynamics of the Delaunay network such as the birth, death and rewiring process of Delaunay edges.
We believe that our method becomes a common tool to understand collective behavior of team sports.

\section*{Acknowledgements}
The authors are very grateful to DataStadium Inc., Japan for providing the player tracking data. 
This work was partially supported by the Data Centric Science Research Commons Project of the Research Organization of Information and Systems, Japan.

\bibliography{./reference} 
\end{document}